\documentclass[aps,prl,superscriptaddress,showpacs,twocolumn]{revtex4-2}
\usepackage[left=1.9cm,right=1.9cm,bottom=1.9cm,top=1.9cm,ignoreall]{geometry}
\usepackage[utf8]{inputenc}
\usepackage{subcaption}
\usepackage{graphicx}
\usepackage[skip=0pt, indent=20pt]{parskip}
\usepackage{amsmath}
\usepackage{amssymb}
\usepackage{xspace}
\usepackage{bm}
\usepackage{color}
\usepackage[squaren,Gray]{SIunits}
\usepackage{empheq}
\usepackage{braket}
\usepackage{physics}
\usepackage{nameref}
\usepackage{multirow}
\usepackage{esvect}
\usepackage{enumerate}
\usepackage[normalem]{ulem}
\usepackage[hyperindex=true]{hyperref}
\usepackage{xcolor}
\hypersetup{linktocpage,colorlinks=true,citecolor=blue,linkcolor=blue}
\usepackage{natbib}
\makeatletter
\let\revtex@section\section
\renewcommand{\section}{%
  \@ifstar
    {\@dblarg\revtex@star@section}
    {\@dblarg\revtex@normal@section}}

\def\revtex@normal@section[#1]#2{%
  \gdef\@currentlabelname{#2}%
  \revtex@section[#1]{#2}%
}
\def\revtex@star@section[#1]#2{%
  \gdef\@currentlabelname{#2}%
  \revtex@section*{#2}%
}
\makeatother

\begin{document}

\title{Deep-Learning-Designed AlGaAs Interface Linking Trapped Ions to Telecom Quantum Networks}

\author{I.P. De Simeone}
\affiliation{Université Paris Cité, CNRS, Laboratoire Matériaux et Phénomènes Quantiques, 75013 Paris, France}

\author{G. Maltese}
\thanks{This work was conducted independently and does not represent the views of Microsoft}
\affiliation{Microsoft Azure Fiber, Romsey, United Kingdom}

\author{V. Cambier}
\affiliation{Université Paris Cité, CNRS, Laboratoire Matériaux et Phénomènes Quantiques, 75013 Paris, France}

\author{J-P. Likforman}
\affiliation{Université Paris Cité, CNRS, Laboratoire Matériaux et Phénomènes Quantiques, 75013 Paris, France}

\author{M. Ravaro}
\affiliation{Université Paris Cité, CNRS, Laboratoire Matériaux et Phénomènes Quantiques, 75013 Paris, France}

\author{L. Guidoni}
\affiliation{Université Paris Cité, CNRS, Laboratoire Matériaux et Phénomènes Quantiques, 75013 Paris, France}

\author{F. Baboux}
\affiliation{Université Paris Cité, CNRS, Laboratoire Matériaux et Phénomènes Quantiques, 75013 Paris, France}

\author{S. Ducci}
\thanks{Corresp. author: sara.ducci@u-paris.fr}
\affiliation{Université Paris Cité, CNRS, Laboratoire Matériaux et Phénomènes Quantiques, 75013 Paris, France}

\begin{abstract}

    The realization of a scalable quantum internet requires efficient light–matter interfaces that map stationary qubits onto photonic carriers for long-distance transmission. A central challenge is the generation of entangled photons simultaneously compatible with single-emitter transitions and low-loss telecom fiber infrastructure. Spontaneous parametric down-conversion in integrated photonic platforms offers a promising route toward this goal. Among available material systems, AlGaAs is particularly attractive due to its large second-order nonlinearity and strong potential for monolithic integration. However, engineering the spectral and spatial properties of the generated quantum states requires the simultaneous optimization of numerous geometric and material parameters, a task remaining computationally demanding for conventional numerical approaches.
    To address this challenge and enable rapid and high-fidelity modeling of complex nonlinear photonic devices, we develop an inverse-design framework based on neural network surrogate models. Using this readily extendable method, we design a transversely pumped AlGaAs waveguide microcavity that produces polarization-entangled photon pairs in distinct spatial modes and frequency channels, one at 1092 nm, resonant with a $^{88}\text{Sr}^{+}$ transition, and the other at 1550 nm in the telecom C-band. This device establishes a direct photonic interface between trapped-ion qubits and long-haul fiber networks, providing a scalable pathway toward hybrid quantum network architectures.

\end{abstract}

\maketitle

\section*{Introduction}

Efficient quantum interfaces between stationary matter qubits and flying photonic qubits are a central prerequisite for the realization of large-scale quantum networks \cite{wei_towards_2022}. Over the past decade, remarkable progress has been achieved in entangling remote quantum nodes via photonic channels using leading matter platforms, including neutral atoms \cite{van_leent_entangling_2022}, quantum dots \cite{zhai_low-noise_2020}, nitrogen-vacancy centers \cite{nemoto_photonic_2016, pompili_realization_2021}, and trapped ions \cite{krutyanskiy_prl_2023, stephenson_high-rate_2020}. In most demonstrations, the emitted photons inherit the native optical transition wavelength of the matter qubit. While this approach is well-suited for local or metropolitan-scale distribution—and, in favorable cases, can be extended to moderate distances—it is fundamentally limited for long-haul quantum communication.
Long-distance transmission through optical fiber can only be efficiently achieved within the low-loss telecom window, in particular the C-band centered at 1550 nm, where attenuation reaches its global minimum. Consequently, a severe spectral mismatch arises between typical matter-qubit transitions—often located in the visible or near-ultraviolet—and the telecom band used by quantum photonic infrastructure. Overcoming this mismatch represents one of the key technological bottlenecks for hybrid and scalable “universal” quantum internet architectures.

Among the available matter platforms, trapped ions stand out for their exceptionally long coherence times and two-qubit gate fidelities exceeding $99.99 \%$ \cite{hughes2025}. Microfabricated ion-chip devices further combine high-fidelity quantum control with compatibility with advanced processing technologies, positioning them among the leading candidates for scalable quantum network nodes. Moreover, recent advances have demonstrated the integration of photonic components directly within ion-trap platforms, including on-chip laser routing and photon collection via grating couplers \cite{knollmann_collection_2025,knollmann_integrated_2024}, marking important steps toward fully integrated quantum network nodes. Nevertheless, efficient long-distance transmission of quantum information from trapped ions remains challenging because their dipole-allowed transitions typically lie far from the telecom band. Quantum frequency conversion (QFC) provides a possible route to bridge this spectral gap. Single- and multi-stage QFC schemes based on bulk nonlinear crystals have successfully interfaced visible-wavelength ion photons with telecom carriers \cite{bock2018, krutyanskiy2023}. In parallel, cavity-enhanced photon collection has significantly increased ion–photon entanglement rates \cite{krutyanskiy_prl_2023}. Despite these advances, frequency-conversion approaches are affected by technological complexity, limited scalability, and may suffer from added noise due to nonlinear conversion processes \cite{cui_metropolitan-scale_2025, baumgart_entanglement_2025}. These limitations motivate the exploration of alternative strategies that directly generate telecom-compatible quantum states without post-emission frequency conversion.

Here we propose a conceptually different approach: an integrated photonic device engineered to generate non-degenerate polarization-entangled photon pairs, with one photon resonant with an ionic transition and the other emitted directly in the telecom C-band. By performing entanglement swapping between the ion-emitted photon and the spectrally matched photon from the chip-based source, ion–telecom-photon entanglement is established. Crucially, the Bell-state measurement used for swapping heralds the successful generation of remote ion–telecom entanglement, enabling conditional network operation and resource-efficient scaling. This architecture bypasses the need for external quantum frequency conversion while preserving compatibility with fiber-based infrastructure.

Integrated quantum photonics provides a compelling technological foundation for this approach, enabling the combination of strong optical nonlinearities, long-term phase stability, and scalability enabled by mature fabrication techniques. Among available material platforms, aluminum gallium arsenide (AlGaAs) is particularly attractive due to its large second-order nonlinear susceptibility, broad transparency window spanning the telecom band, strong electro-optic response, and compatibility with monolithic laser integration \cite{Baboux2023}. AlGaAs waveguides have already demonstrated efficient spontaneous parametric down-conversion (SPDC), enabling compact and low-noise entangled-photon sources suitable for large-scale deployment \cite{appas_flexible_2021}.
However, designing multilayer AlGaAs waveguides for highly non-degenerate SPDC is a complex task, as efficient photon-pair generation demands the simultaneous fulfillment of multiple constraints that often impose competing requirements on the structural parameters.
Conventional grid-based parameter sweeps rapidly become computationally prohibitive and restrict exploration to narrow subspaces of the full design domain.

To overcome these limitations, we introduce a neural-network–based inverse-design framework enabling fully differentiable optimization of nonlinear multilayer AlGaAs waveguides. We focus specifically on an integrated source engineered to emit photon pairs in which one photon lies in the telecom C-band and the other is resonant with the 1092 nm dipole-allowed $5P_{1/2} \to 4D_{3/2}$ transition of the $^{88}\text{Sr}^{+}$ ion (linewidth $\approx 20 MHz$). This choice of ion species is motivated by its favorable spectral properties: spontaneous emission at 1092 nm lies significantly closer to the telecom band than the principal dipole-allowed transitions of other widely used trapped-ion species. Indeed, this transition has already enabled kilometer-scale ion–photon entanglement via direct fiber injection \cite{zalewski2026}. In our scheme, the reduced spectral separation between 1092 nm and 1550 nm substantially relaxes phase-matching constraints for non-degenerate SPDC, enabling efficient chip-scale implementation.

The resulting device—realized as a transversely pumped AlGaAs waveguide microcavity—constitutes a compact and scalable source of entangled photons tailored for ion–photon interfacing. Beyond its immediate relevance for hybrid ion-based quantum networks, our work highlights the broader potential of differentiable, machine-learning–assisted photonic design for complex nonlinear integrated devices operating under stringent multiparameter constraints.

\section*{Results}

\begin{figure*}[bt]
    \centering        \includegraphics[width=\textwidth]{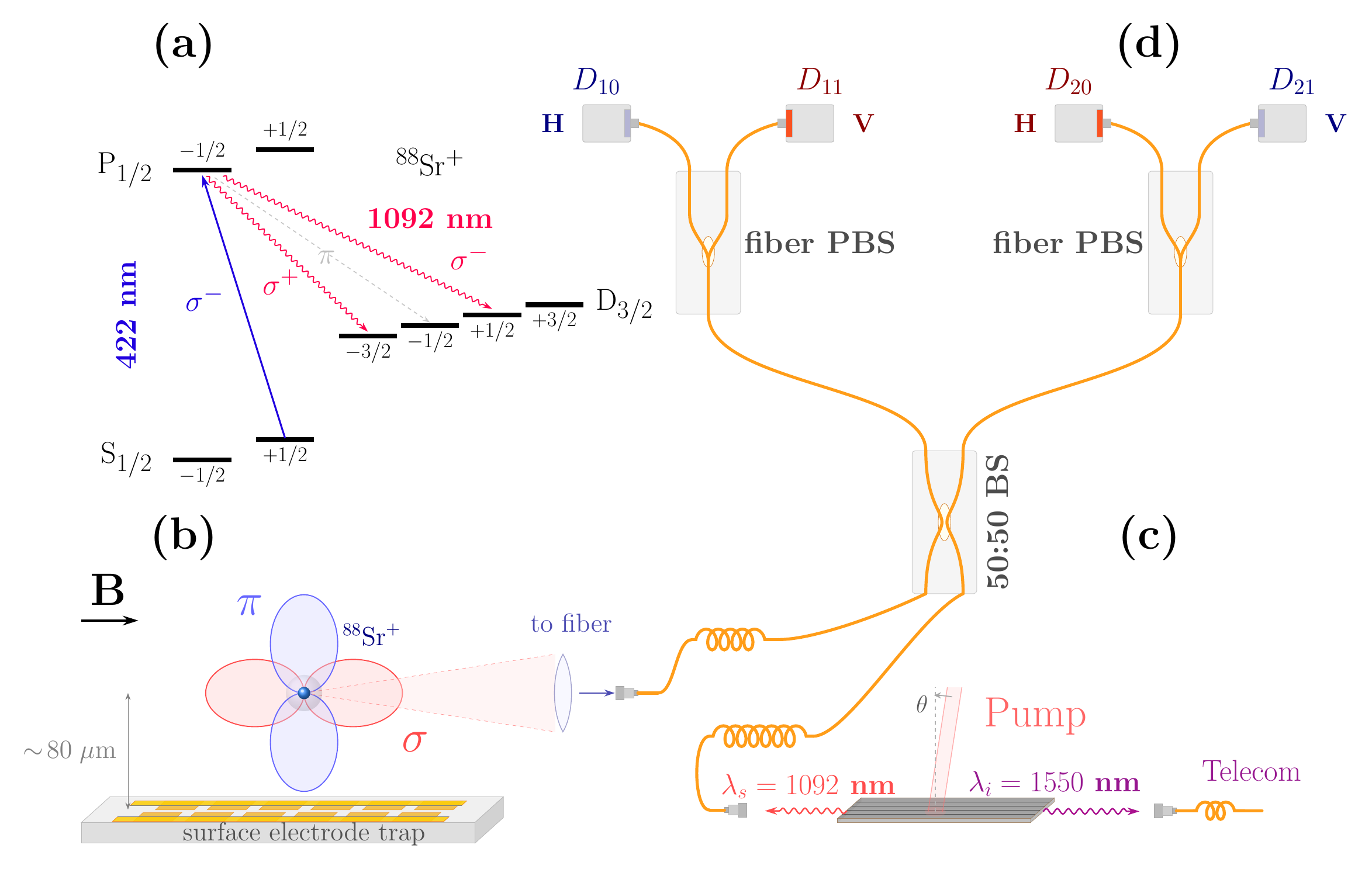}
    \caption{\textbf{Experimental setup of the quantum interface between non-degenerate SPDC source and trapped strontium ions}. \textbf{(a)} Simplified level scheme of the $^{88}Sr^+$ ion. A 422 nm $\sigma^-$ - polarized laser pulse coherently drives the transition from $S_{1/2}$ to $P_{1/2}$. Ion photon entanglement is generated through the 1092 nm spontaneous decay. \textbf{(b)} Ion-trap apparatus. A single ion confined above a surface trap spontaneously decays. The $\pi$-polarized emission is not collected, while the $\sigma$-polarized photons are coupled into optical fibers and routed to the Bell-state measurement setup. \textbf{(c)} Schematic view of the non-degenerate SPDC source. A pump beam impinges on the waveguide surface, generating polarization-entangled photon pairs with one photon at a telecom wavelength and the other matched to 1092 nm, corresponding to the considered ion transition.\textbf{(d)} Bell-state measurement setup. Entanglement swapping is implemented using a 50:50 fiber beam splitter (BS) followed by two fibered polarizing beam splitters (PBS).}

    \label{fig:experimental_scheme_full}
\end{figure*}

The first step of the entanglement generation between the ion electronic state and the photon polarization state relies on selection rules and spontaneous emission (Figure \ref{fig:experimental_scheme_full}a.)
The protocol is the following: we start by coherently transferring the electronic population to the
$ \ket{5p\, ^2 P_{1/2},-\frac{1}{2}} $ Zeeman sublevel.
If a spontaneous photon is then emitted at $\lambda_{ion} = 1092$ nm ( $5P_{1/2} \to 4D_{3/2}$ transition, probability 0.056 \cite{likforman_precision_2016}), the emitted photon polarization state and the final electronic state (Zeeman sublevel) of the ion become entangled.
If the collection optics is aligned along the quantification axis defined by an external magnetic field $\vec{B}$, the shapes of spontaneous emission diagrams (Figure 1b) allow us to consider only circularly polarized photons, implying an entangled state of the form:

\begin{equation}\label{eq:ion-state}
    \frac{\sqrt{3}}{2}\ket{D_{3/2},-\frac{3}{2}}_{ion}\ket{\sigma^-}_{phot} + \frac{1}{2}\ket{D_{3/2},\frac{1}{2}}_{ion}\ket{\sigma^+}_{phot}
\end{equation}
This state is non-maximally entangled because the Clebsch-Gordan coefficients for the two decay channels are not equal. However, a maximally entangled state can be recovered at the expense of the entanglement rate.
A quarter-wave plate can be employed to transform the circular polarizations of the emitted photons into the more standard $HV$ basis.

As explained above, we interface the trapped ion with the C-band telecom infrastructure using a photonic device that emits nondegenerate signal/idler polarization-entangled photon pairs (see Figure \ref{fig:experimental_scheme_full}c).
\begin{equation}
    \frac{1}{\sqrt{2}}(\ket{H}_s\ket{V}_i + \ket{V}_s\ket{H}_i)
\end{equation}
In this scheme, the signal photon must be spectrally matched to the ion emission at $\lambda_{s} = 1092$ nm, while the idler photon must lie in the telecom C band at $\lambda_{i} \simeq 1550$ nm.

A Bell-state measurement (BSM) performed between the signal photon and the photon emitted by the $^{88}\text{Sr}^{+}$ ion (Figure \ref{fig:experimental_scheme_full}d) projects the ion and the idler photon into an entangled state (entanglement swapping) \cite{Zeilinger-swapping}. The idler photon thus becomes a flying qubit, enabling long-distance entanglement distribution through a quantum network.

The semiconductor entangled photon source is implemented as a microcavity ridge waveguide based on a counterpropagating phase-matching scheme \cite{orieux_efficient_2011, orieux_direct_2013}. In this configuration, the device operates under transverse pumping (see Fig. \ref{fig:counterprop-state}a) The pump beam impinges on the top surface of the waveguide at an angle $\theta$ and undergoes spontaneous parametric down-conversion into two orthogonally polarized photons that propagate in opposite directions along the waveguide axis. This geometry provides several key advantages for the targeted ion–photon interface.
(i) The counterpropagating phase-matching condition yields photon pairs that are at least two orders of magnitude spectrally narrower than those generated in co-propagating geometries, thereby facilitating efficient coupling to photons emitted via narrow atomic transitions.
(ii) The signal and idler photons are separated into distinct spatial modes at the generation stage, eliminating the need for additional separation optics.
(iii) Polarization-entangled states can be generated intrinsically through appropriate pump-beam engineering, eliminating the need for walk-off compensation \cite{orieux_direct_2013}.
(iv) The emission wavelengths can be tuned by adjusting the pump incidence angle, offering a convenient and flexible means to match the ion transition frequency. This tunability is illustrated later through the device’s tuning curves.

\begin{figure*}[bt]
    \centering
    \includegraphics[width=\textwidth]{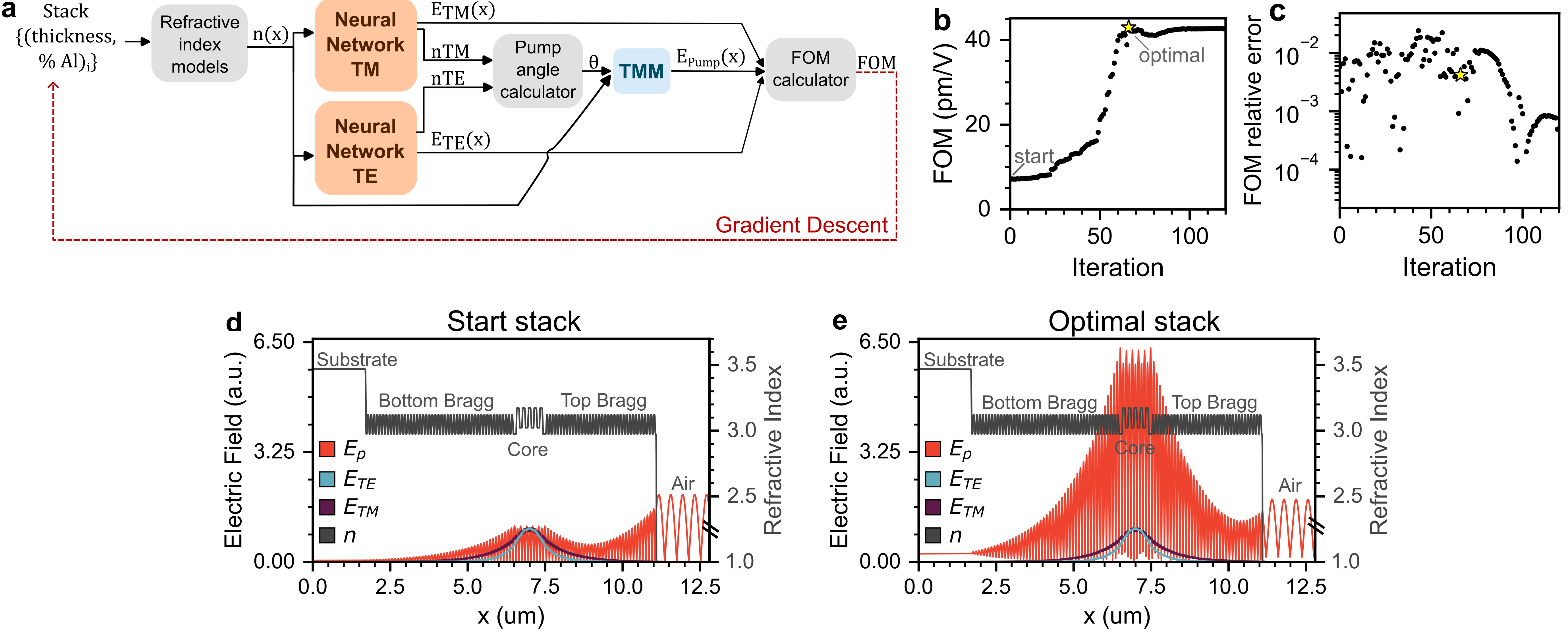}
    \caption{Ion-photon interface design: (a) Workflow for source optimization (see text). (b) Evolution of the figure of merit (FOM) throughout the optimization process. The yellow star indicates the optimal configuration corresponding to the highest FOM. (c) Relative discrepancy between our neural-network-based surrogate model and a conventional transfer-matrix-method. (d)-(e): Comparison between the initial (d) and optimized (e) designs: the black curve shows the refractive-index profile of the multilayer stack. The blue and purple curves correspond to the spatial distributions of the fundamental TE and TM guided modes of the generated photons, respectively, while the red curve represents the transverse profile of the pump field.}
    \label{fig:counterpropa-opt}
\end{figure*}

\begin{figure*}[t]
    \centering
    \includegraphics[scale=0.6,width=\textwidth]{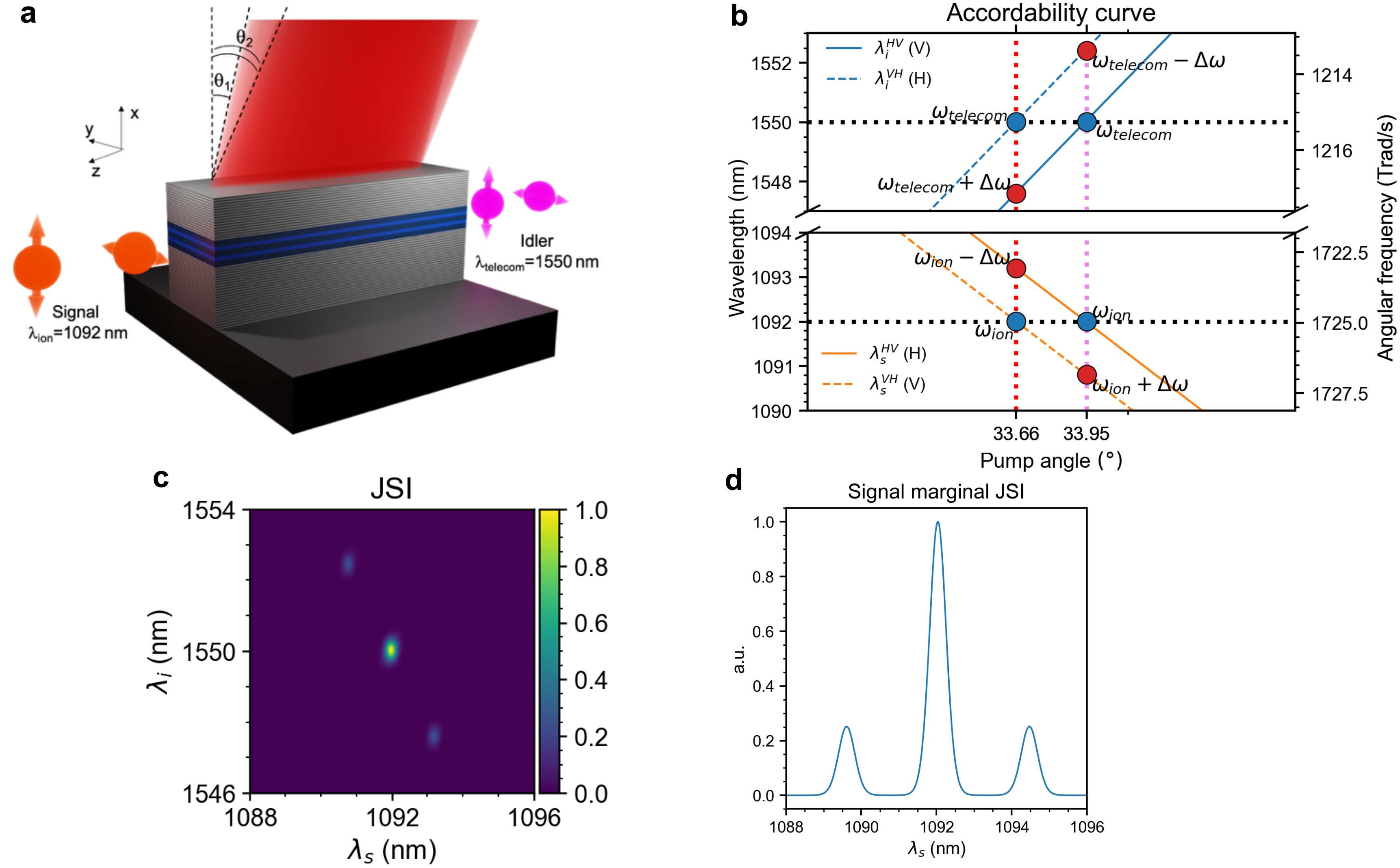}
    \caption{Counterpropagating ion-photon interface characteristics. (a) Schematic of the device and the pump-incidence geometry for the generation of non-degenerate polarization-entangled photons. (b) Tunability curves showing the dependence of the photon pair wavelengths on the pumping angle. (c) Joint Spectral Intensity (JSI) of the biphoton state emitted by the device when pumped simultaneously at $\theta_1 \approx 33.66^\circ$ and $\theta_2 \approx 33.95^\circ$. (d) Marginal Joint Spectral Intensity for the signal photon, highlighting the peak at the ion transition wavelength.}
    \label{fig:counterprop-state}
\end{figure*}

The device’s broad operational flexibility comes at the cost of design complexity. As will be illustrated by the results presented in Fig. \ref{fig:counterpropa-opt}e, the microcavity is formed by a vertically engineered stack of epitaxially grown AlGaAs layers with precisely controlled aluminum compositions. The top and bottom Bragg reflectors establish the vertical microcavity, providing resonant enhancement of the pump field while simultaneously confining the generated photon pairs within the waveguide core.
The core region consists of a multilayer sequence with alternating high- and low-aluminum fractions, implementing quasi-phase matching (QPM) along the growth direction.
For the target wavelengths $\lambda_{s} = 1092$ nm and $\lambda_{i} = 1550$ nm, energy conservation sets the pump wavelength to $\lambda_{p} = 640$ nm. This requirement imposes a stringent constraint on the material composition: to suppress absorption at the pump wavelength, all AlGaAs layers must contain at least 50$\%$ aluminum, ensuring a sufficiently wide bandgap \cite{vurgaftman_2001}.

The simultaneous optimization of the pump‑cavity resonance, material transparency, mode confinement, and QPM conditions is far from trivial. A traditional strategy is to rely on numerical simulations combined with grid‑based parameter searches. However, the parameter space spans numerous degrees of freedom,  defined by the thickness and aluminum composition of each AlGaAs layer, making grid-search exploration computationally prohibitive.
To efficiently explore the high‑dimensional design space, we employ an inverse‑design approach \cite{molesky2018inverse,khaireh-walieh_newcomers_2023}.
This strategy requires a fully differentiable modeling pipeline, incompatible with standard electromagnetic solvers. 
Inspired by a recent work \cite{lim2022maxwellnet}, we train neural‑network surrogate models to predict both the effective indices and the transverse electric‑field profiles of the guided TE and TM infrared modes directly from the multilayer refractive‑index profile. These models, combined with a differentiable implementation of the transfer‑matrix method (TMM) \cite{chilwell_thin-films_1984}, provide the interacting fields required to evaluate the relevant figure of merit (FOM) within the inverse‑design framework (in this case, the nonlinear overlap integral between the interacting fields).

Figure \ref{fig:counterpropa-opt}a summarizes the design workflow.
The initial epitaxial stack - defined by a set of layer thicknesses and aluminum concentrations -  is first translated into a refractive-index profile using established dispersion models \cite{afromowitz_refractive_1974, gehrsitz_refractive_2000}. To ensure differentiability within the optimization framework, the nominal step‑index profile is implemented with smooth transitions rather than discontinuous steps between layers. The refractive‑index profile is then fed as input to two neural‑network surrogate models, trained to predict the guided modes' properties: the TE electric field distribution $E_{TE}(x)$ and the TM electric displacement field $D_{TM}(x)$, together with their corresponding effective refractive indices $n_{TE}$ and $n_{TM}$. For the TM polarization, the model is trained to predict $D_{TM}$ rather than $E_{TM}$
since the continuity of the displacement field across material interfaces improves numerical stability and accelerates convergence during training. The TM electric field  $E_{TM}$ is subsequently reconstructed from $D_{TM}$.
The modes' effective indices are used to determine the pump incidence angle $\theta$ that satisfies the longitudinal phase‑matching condition
via
$\sin{\theta} ={\lambda_p}\left( \frac{n_{TE}}{\lambda_{TE}} - \frac{n_{TM}}{\lambda_{TM}}\right) $.
Because the pump field is not a guided eigenmode of the waveguide, but rather an obliquely incident wave impinging onto the waveguide, its transverse electric-field profile $E_p(x)$ is computed using a fully differentiable transfer-matrix-method (TMM) solver developed specifically for this work. The resulting pump profile is combined with the TE and TM guided-mode profiles to evaluate the FOM. Owing to the fully differentiable nature of the entire computational pipeline, gradients of the FOM with respect to the design parameters—namely, the layer thicknesses and aluminum compositions—are efficiently obtained via backpropagation and subsequently used in a gradient-based optimization routine.

The optimization objective is to maximize the SPDC conversion efficiency, which scales with the squared magnitude of the nonlinear overlap integral $|\Gamma|$. This quantity is determined by the spatial overlap of the three interacting fields along the epitaxial growth axis $x$, and is given by:
\begin{gather}
    \Gamma = \int dx \; \chi^{(2)}(x) E_p(x) E_{TE}(x) E_{TM}(x) \label{eq:overlap-integral}
\end{gather}

where $\chi^{(2)}(x)$ is the second-order nonlinear susceptibility of the AlGaAs stack. Since $|\Gamma|$ itself is  a smooth function of the optimization parameters, it is adopted as the figure of merit (FOM). The maximization is then carried out iteratively using the Adam optimizer, enabling efficient convergence within the high-dimensional design space \cite{kingma2017adammethodstochasticoptimization}.
Notably, the quasi‑phase‑matching condition along the growth direction does not need to be imposed explicitly: it is inherently captured by the spatial modulation of $\chi^{(2)}(x)$ in Eq. \eqref{eq:overlap-integral}. Designs that do not satisfy vertical QPM naturally yield smaller overlap values, i.e. FOMs, and are therefore disfavored during optimization.


Figure \ref{fig:counterpropa-opt}b presents the evolution of FOM throughout the optimization process.
As expected for a gradient‑based optimization scheme, it increases monotonically with each iteration until it eventually reaches a plateau.
The initial structure yields a FOM of $\approx 7$ pm/V, while the optimized design achieves $\approx 43$ pm/V. Given the quadratic dependence of the conversion efficiency on the FOM, this sixfold increase in $|\Gamma|$ translates into an overall $\sim$36-fold enhancement in the conversion efficiency. The small overshoot observed before convergence is attributed to the adaptive step size of the Adam optimizer, which can temporarily drive the parameters beyond the local optimum before settling into the stationary regime. To assess the accuracy and robustness of the proposed framework, the FOM is evaluated at each iteration using both neural-network-based surrogate models and a conventional, non-differentiable transfer-matrix method (TMM) calculation. The relative discrepancy between the two approaches is shown in Fig.~\ref{fig:counterpropa-opt}c. The excellent agreement throughout the entire optimization trajectory confirms that the differentiable framework accurately reproduces the underlying electromagnetic response.

The epitaxial structure of the final optimized device is summarized in Table~\ref{tab:interface-epitax-struct}. A direct comparison between the initial and optimized stacks is shown in Figure \ref{fig:counterpropa-opt}d-e. The black curve shows the refractive-index profile of the multilayer stack, extending from the substrate to air. The blue and purple curves correspond to the spatial distributions of the fundamental TE and TM guided modes of the generated photons, respectively, while the red curve represents the transverse profile of the pump field.

In the initial structure, the microcavity is not resonant at the pump wavelength, resulting in strong reflection at the cavity interface. As a result, only a small fraction of the pump field penetrates the nonlinear core, leading to a conversion efficiency on the order of $5 \times 10^{-13} \, \text{pairs}/\text{pump photon}$. In contrast, the optimized design is resonant at the pump wavelength - which substantially enhances the intracavity field intensity within the core region -, while simultaneously satisfying the multiple design constraints. This resonant enhancement increases the nonlinear interaction strength and yields a two-order-of-magnitude improvement in conversion efficiency, reaching approximately $2 \times 10^{-11} \, \text{pairs}/\text{pump photon}$. Such performance is equivalent to that reported for degenerate photon-pair sources based on similar microcavity geometries \cite{orieux_efficient_2011}.

We now describe how the proposed device enables the direct generation of polarization-entangled photon pairs at the two target wavelengths by appropriately choosing the pump geometry. When the pump beam impinges on the top surface of the waveguide at an angle $\theta$, the counterpropagating configuration of the signal and idler modes supports two simultaneous, phase-matched SPDC processes. In the first process, the signal photon, propagating in the positive $z$ direction (see Fig.\ref{fig:counterprop-state}a), is TE-polarized (denoted H), while the idler photon, propagating in the negative $z$ direction, is TM-polarized (denoted V). This is referred to as the HV process. In the second process, the polarizations are reversed: the signal is TM-polarized (V) and the idler is TE-polarized (H), corresponding to the VH process.

Energy and momentum conservation uniquely determine the frequencies of the generated photons for a given pump incidence angle $\theta$. Under ideal phase-matching conditions, each pump angle therefore supports two distinct photon-pair generation channels with different frequency combinations, arising from the slight birefringence between the two orthogonally polarized modes \cite{orieux_direct_2013}. As illustrated in Fig.\ref{fig:counterprop-state}b, simultaneous excitation of the structure at $\theta_1 \approx 33.66^\circ$ and $\theta_2 \approx 33.95^\circ$ activates four phase-matched processes overall, resulting in the concurrent generation of four types of photon pairs. The quantity $\Delta \omega$ shown in the figure represents the shift in angular frequency between the two photon-pair generation channels arising from birefringence.

By spectrally filtering out the two pairs that do not contain a photon resonant with the targeted ion transition wavelength, only two indistinguishable generation pathways remain. Because these processes are simultaneously phase-matched and share the same spatial and spectral characteristics after filtering, their probability amplitudes add coherently. The emitted two-photon state can therefore be written as
\begin{equation}
    \ket{\omega_{ion}}_s\ket{\omega_{telecom}}_i\frac{1}{\sqrt{2}}\Big[ \ket{H}_s\ket{V}_i + \ket{V}_s\ket{H}_i\Big]
\end{equation}

with $\omega= 2 \pi c/\lambda$.
This corresponds to a polarization-entangled Bell state, directly generated at the two wavelengths of interest (see Fig. \ref{fig:counterprop-state}a) .
Generalizing the previous discussion beyond the ideal case of perfect phase matching, both the HV and VH processes contribute with angle- and frequency-dependent probability amplitudes determined by the corresponding phase-matching functions. In this more general scenario, the generated two-photon state is described by a coherent superposition weighted by the joint spectral amplitudes (JSA) of the two processes:

\begin{equation}\label{eq:tot-jsa}\begin{split}
        \ket{\psi} = \iint d\omega_1 d\omega_2 \Big[ & \phi_{HV}(\omega_1,\omega_2)\hat{a}^{\dagger}_{H,s}(\omega_1)\hat{a}^{\dagger}_{V,i}(\omega_2) +             \\
                                                     & \phi_{VH}(\omega_1,\omega_2)\hat{a}^{\dagger}_{V,s}(\omega_1)\hat{a}^{\dagger}_{H,i}(\omega_2) \Big] \ket{0}
    \end{split}
\end{equation}
where $\phi_{HV}(\omega_s,\omega_i)$ and $\phi_{VH}(\omega_s,\omega_i)$ are the joint spectral amplitudes associated with the two phase-matching channels \cite{Guillaume-toolbox}. These functions incorporate the pump spectral envelope as well as the finite phase-matching bandwidth of the structure.

Figure \ref{fig:counterprop-state}c shows the calculated joint spectral intensity - the modulus squared of the JSA - of the biphoton state generated when the device is simultaneously pumped at $\theta_1 \approx 33.66^\circ$ and $\theta_2 \approx 33.95^\circ$. The central overlapping region corresponds to the two target, indistinguishable HV and VH processes that produce the polarization-entangled state. The side lobes arise from the additional phase-matched processes activated at these pump angles; these contributions are spectrally separated and can be suppressed using appropriate filtering. Figure \ref{fig:counterprop-state}d shows the Marginal Joint Spectral Intensity for the signal photon featuring a peak at the ion transition wavelength $\lambda_{\rm ion} = 1092$ nm.

\begin{table}[bht]
    \centering
    \begin{tabular}{|c|c|c|c|}
        \hline
        Number of periods   & Role                  & Al content (\%) & Thickness (nm) \\
        \hline
                            & Substrate             & 0               &                \\
        \hline
        \multirow{2}{*}{48} & Bragg                 & 90              & 48.4           \\
        \cline{3-4}
                            & Bottom                & 60              & 58.6           \\
        \hline
        1                   & Buffer                & 90              & 120.7          \\
        \hline
        \multirow{2}{*}{4}  & \multirow{2}{*}{Core} & 50              & 91.7           \\
        \cline{3-4}
                            &                       & 80              & 96.4           \\
        \hline
        1                   & Buffer                & 90              & 120.7          \\
        \hline
        \multirow{2}{*}{36} & Bragg                 & 60              & 58.6           \\
        \cline{3-4}
                            & Top                   & 90              & 48.4           \\
        \hline
    \end{tabular}
    \caption{Ion-photon interface epitaxial structure}
    \label{tab:interface-epitax-struct}
\end{table}

\section*{Discussion}



\subsection*{Performances}

We begin our discussion by estimating the heralded ion–photon entanglement rate achievable with the integrated photon-pair source described above.
On the ion side, we assume state-of-the-art free-space photon collection with a numerical aperture (NA) of 0.6 and a fiber-coupling efficiency of 70 \%. We further account for the branching ratio of the $P_{1/2} \rightarrow D_{3/2}$ transition (0.056), an experimental attempt rate of 1 MHz, and a photon-detection time window of 50 ns.
On the source side, we consider synchronous pumping with a nanosecond pulsed laser delivering 60 mW average power at a repetition rate of 1 MHz. We also include the spectral overlap between the photon-pair source and the ion transition, corresponding to an estimated mismatch factor of 50.
Combining these contributions yields an expected heralded ion–photon entanglement rate on the order of two events per minute.


Several strategies may be envisaged to increase this rate, starting with improvements on the ion side. Parallelization of entanglement attempts—e.g., through multiplexing within an ion string \cite{cui_metropolitan-scale_2025}—can enhance the overall rate approximately linearly with the number of simultaneously addressed ions. The photon-collection efficiency may also be significantly improved by coupling the ions to optical microcavities \cite{krutyanskiy_prl_2023}, thereby increasing the emission into the desired spatial mode. In addition, employing alternative photonic degrees of freedom, such as time-bin encoding rather than polarization, can relax constraints on the collection optics and improve robustness against optical imperfections. Finally, the development of ion traps with integrated collection waveguides represents a promising direction. Such architectures may ultimately outperform free-space collection in efficiency, while also offering improved scalability and integration prospects for large-scale quantum-network nodes.

Concerning the AlGaAs source, several strategies can be pursued to further enhance its performance. First, spectral matching between the photon bandwidth and the targeted ion transition can be improved by depositing high-reflectivity coatings on the waveguide facets at 1092 nm. Standard techniques, such as ion-beam-assisted deposition of $SiO_2$/$TiO_2$ Bragg mirrors, can realistically achieve reflectivities in the 99–99.9\% range. Such reflectivities would result in cavity resonance linewidths on the order of 24-240 MHz, significantly improving spectral overlap with the ionic transition.
A second promising approach to increase the entangled-photon emission rate is spatial multiplexing \cite{Meyer-Scott2020}. Owing to the maturity of the AlGaAs fabrication platform, a large number of nominally identical waveguides can be integrated on a single chip. In this configuration, the overall source brightness scales linearly with the number of waveguides, while the higher-order multi-pair emission probability remains unchanged for each individual source. 
Moreover, the recent demonstration of on-chip photon-pair generation combined with integrated polarization or 50/50 beam splitters in AlGaAs waveguides \cite{Appas2023,Belhassen2018} paves the way toward the realization of an on-chip Bell-state measurement. Integrating these functionalities within a single photonic platform would reduce coupling losses and experimental complexity, while improving phase stability and scalability. Ultimately, this level of integration would further advance the ion–photon interface presented here toward a compact, robust, and scalable quantum-network node.

\subsection*{Inverse design framework generalization}

Finally, beyond the specific use case explored in this article, the developed inverse design framework can be extended and adapted in multiple directions. As demonstrated, this approach — based on deep learning and surrogate modeling — offers substantial flexibility and computational efficiency. It enables rapid exploration of large parameter spaces while reconciling multiple competing optical and material constraints. By directly optimizing figures of merit expressed in terms of guided-mode profiles, effective indices, and structural parameters, the method provides a systematic route toward highly integrated and application-specific nonlinear photonic devices.
The differentiable framework presented here can be readily generalized by training the neural networks on alternative epitaxial structures, thereby extending its applicability to a broad class of photonic devices, including VCSELs, photodiodes, and heterostructure laser diodes. With suitable refractive-index models, the approach can also be adapted to other material platforms such as InGaAs, InGaP, and InGaAsP.
If the considered epitaxial stacks differ substantially from those represented in the original training dataset, the surrogate models can be retrained from scratch to ensure predictive accuracy. For more moderate deviations, the models can instead be fine-tuned during the optimization process, restoring accuracy while maintaining a limited additional computational overhead.
Beyond these immediate generalizations, several promising extensions can be envisioned. First, the surrogate models could be trained to predict mode profiles and effective indices over a frequency range rather than at a single design wavelength. Incorporating spectral dependence directly into the optimization workflow would enable the treatment of dispersion-related effects and facilitate the design of broadband devices whose performance critically depends on chromatic dispersion.
A second important development would be the explicit inclusion of fabrication tolerances within the optimization objective. By accounting for realistic process variations during training or optimization, the framework could identify intrinsically robust device geometries, thereby improving manufacturability and yield.
In the longer term, a particularly compelling direction is the end-to-end optimization of complete quantum information protocols. This would encompass not only photon-pair generation, but also their subsequent manipulation and measurement. Since quantum-mechanical processes can be formulated as linear-algebra operations that are naturally differentiable, they can, in principle, be integrated seamlessly into the same computational graph. Such an approach would enable holistic, system-level optimization of complex quantum-optical architectures within a unified differentiable framework.

\section*{Materials and methods}
\subsection{Optimization loop}
\paragraph{Field normalization convention}

The signal and idler mode profiles correspond to the fundamental TE and TM waveguide eigenmodes. They have units of $\mathrm{nm}^{-1/2}$ and are normalized to carry unit power across the entire epitaxial structure:
\begin{equation}
    \int dx  E_m^2(x) = 1, \qquad m = \mathrm{TE,TM}.
\end{equation}
The pump field is modeled as a plane wave propagating in the $x$–$z$ plane. Its electric field is written as
\begin{equation}
    E_p(\vec{r}) = A_p(z)\big[ \Phi_p^+(x) + \Phi_p^-(x) \big] e^{i k_p \sin\theta z}
\end{equation}
Here, $A_p(z)$ (in V/nm) denotes the slowly varying field amplitude along the propagation direction. The dimensionless functions $\Phi_p^+(x)$ and $\Phi_p^-(x)$ represent the transverse field components associated with waves propagating toward the substrate and toward the air, respectively.
The pump field is normalized such that
\begin{equation}
    \left|\Phi_p^+(x_0)\right| = 1
\end{equation}
where $x_0$ denotes the waveguide–air interface.

\paragraph{Choice of the initial structure}

The choice of the initial structure is critical for guiding the optimization process and preventing convergence to sub-optimal local minima. Based on physical considerations, we first define the aluminum concentration profile. To enhance both Bragg-mirror reflectivity and quasi-phase-matching (QPM) efficiency, we impose a large refractive-index contrast by alternating the minimum and maximum aluminum fractions allowed within the design space. Furthermore, the average aluminum concentration in the core region is selected to be lower than that of the surrounding Bragg mirrors and buffer layers, ensuring effective confinement of the infrared (IR) modes. These physically motivated trends are preserved in the final optimized structure (Table \ref{tab:interface-epitax-struct}).
The initial layer thicknesses are chosen following similar physical arguments. For pump confinement, the optimal thickness of an individual Bragg-mirror layer is approximately $\lambda/4$, while the vertical QPM condition requires core-layers thickness close to $\lambda/2$. Although these characteristic values are not expected to be strictly optimal for nonlinear conversion—due to competing optical and phase-matching constraints—the global optimum is anticipated to lie in their vicinity. Accordingly, the Bragg-mirror thicknesses are initialized by random sampling within the range $ \lambda/4 \pm 25 \% $, and the QPM core-layers thicknesses within  $\lambda/2 \pm 25 \%$.
The buffer-layer thickness is independently sampled from a uniform distribution between 50 and 200 nm.

\paragraph{Fine-tuning the neural networks}

To ensure high-fidelity predictions throughout the optimization process, the neural-network surrogate models are periodically fine-tuned. This procedure consists of a short additional training phase—typically on the order of a few tens of epochs—using data generated from the most recently explored epitaxial structure. Training continues until the mean squared error (MSE) loss falls below a threshold of $10^{-6}$.
This adaptive fine-tuning strategy enables the surrogate models to progressively specialize in the region of the design space being under exploration. As a result, predictive accuracy is maintained locally, thereby improving the robustness and overall reliability of the optimization procedure.

\paragraph{Optimization loop hyper-parameters}
The optimization loop employs the Adam algorithm. A systematic search over the hyperparameter space was performed to ensure stable and efficient convergence. The following configuration was found to provide optimal training performance: learning rate $= 5.0 \times 10^{-4}$, $\beta_1 = 0.9$, and $\beta_2 = 0.999$.

\paragraph{Neural network training}
The neural-network training and the optimization loop are implemented in PyTorch \cite{Ansel_PyTorch_2_Faster_2024}. A comprehensive description of the network architectures, training datasets, and optimization procedures will be presented in a future publication.

\section*{Data availability}
Data are available upon reasonable request

\section*{Acknowledgments}
This work was supported by the Plan France 2030 through projects ANR-22-PETQ-0006 and ANR-22-PETQ-0011, and by the European Union's Horizon Europe research and innovation programme under the project Quantum Secure Networks Partnership (QSNP, Grant Agreement No. 101114043).
We acknowledge Matteo Karr-Ducci for Fig. 3 a).

\section*{Conflicts of interest}
The authors declare no conflicts of interest.
\bibliography{bibliography}

\end{document}